\documentclass[10pt,conference]{IEEEtran}
\usepackage{amsmath,amsfonts,amsthm,amssymb}
\usepackage{tikz}
\usetikzlibrary{calc}
\usepackage{cite}

\newtheorem{theorem}{Theorem}

\IEEEoverridecommandlockouts

\begin{document}
\allowdisplaybreaks

\title{Can Feedback Increase the Capacity of the Energy Harvesting Channel?}
\author{\IEEEauthorblockN{Dor Shaviv}%
\IEEEauthorblockA{
EE Dept., Stanford University\\
shaviv@stanford.edu}%
\and%
\IEEEauthorblockN{Ayfer \"{O}zg\"{u}r}%
\IEEEauthorblockA{
EE Dept., Stanford University\\
aozgur@stanford.edu}%
\and%
\IEEEauthorblockN{Haim Permuter}%
\IEEEauthorblockA{
ECE Dept., Ben-Gurion University\\
haimp@bgu.ac.il}%
\thanks{The work of D.~Shaviv and A.~\"{O}zg\"{u}r was partly supported by a Robert Bosch Stanford Graduate Fellowship and the Center for Science of Information (CSoI), an NSF Science and Technology Center, under grant agreement CCF-0939370. The work of H.~Permuter was supported by the Israel Science Foundation (grant no. 684/11) and the ERC starting grant.}
}

\maketitle

\begin{abstract}

We investigate if feedback can increase the capacity of an energy harvesting communication channel where a transmitter powered by an exogenous energy arrival process and equipped with a finite battery communicates to a receiver over a memoryless channel. For a simple special case where the energy arrival process is deterministic and the channel is a BEC, we explicitly compute the feed-forward and feedback capacities and show that feedback can strictly increase the capacity of this channel. Building on this example, we also show that feedback can increase the capacity when the energy arrivals are i.i.d. known noncausally at the transmitter and the receiver.   

\end{abstract}

\section{Introduction}
\label{sec:introduction}

The capacity of the basic energy harvesting communication channel where the transmitter is powered by an exogenous energy arrival process and equipped with a battery of size $B_{\mathrm{max}}$ has been of significant recent interest \cite{ozel2012achieving,ozel2011awgn,tutuncuoglu2013binary,
mao2013capacity,dong2014approximate,tutuncuoglu2014binary,
shaviv2015capacity}. The capacity has been characterized in the two extremal cases $B_{\mathrm{max}}=\infty$ and $B_{\mathrm{max}}=0$ in \cite{ozel2012achieving} and \cite{ozel2011awgn} respectively. When $B_{\mathrm{max}}$ is finite, \cite{shaviv2015capacity} characterizes the capacity as the limit of an $n$-letter mutual information rate under various assumptions on the availability of energy arrival information at the transmitter and/or the receiver, and derives upper and lower bounds which are easier to compute, and which differ by a constant gap. The difficulty in characterizing the capacity in this case lies in the fact that the channel has an input-dependent state with memory which is known at the transmitter but not at the receiver. In this paper, we consider the question of whether feedback can increase the capacity of this peculiar channel.

\begin{figure}[!t]
\centering
\begin{tikzpicture}
	\node[draw,rectangle] (Tx) at (0,0) {Transmitter};
	\node[draw,rectangle] (Channel) at (2.5,0) {Channel};
	\node at (2.5,0.5) {BEC};
	\node[draw,rectangle] (Rx) at (5,0) {Receiver};
	\node[draw,rectangle] (Battery) at (0,1) {Battery};
	\node at (-1.2,1) {$B_{\mathrm{max}}$};
	
	\tikzstyle{every path}=[draw,->]
	\path (Battery) -- (Tx);
	\path (0,1.8) -- node[above,pos=0.1] {$E_t$} (Battery);
	\path (Tx) -- node[above] {$X_t$} (Channel);
	\path (Channel) -- node[above] {$Y_t$} (Rx);
	\path[dashed,style={rounded corners}] 
		(Rx) -- ++(0,-0.7) -| (Tx)
		node[below,pos=0.25] {$Y_{t-1}$};
\end{tikzpicture}
\caption{Energy harvesting channel model.}
\label{fig:sysmodel}
\end{figure}
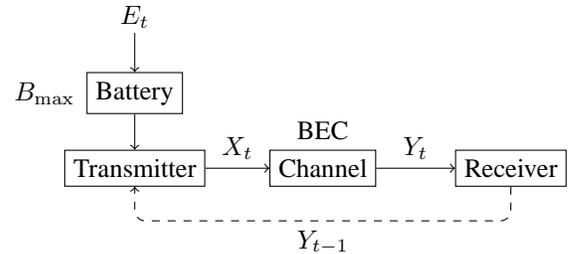

Feedback naturally comes into play in certain applications of energy harvesting networks where the transmitter is powered by RF energy transfer from its corresponding receiver. Such applications include internet of things, where many tiny self-powered sensor nodes may be communicating to a sink node which has access to conventional power, or RFID tags. In such applications it can be natural for the receiver to provide feedback information to the transmitter along with RF energy. We model such a communication scenario with a simple model. We assume that the transmitter has a unit battery which is recharged periodically every two channel uses and the communication occurs over a Binary Erasure Channel (BEC). See Fig.~\ref{fig:sysmodel}. We compute the capacity of this channel with and without causal output feedback from the receiver to the transmitter and show that the feedback capacity is strictly larger than the feed-forward capacity. We then extend our model to the case of i.i.d. Bernoulli battery recharges and show that when the energy arrivals are known noncausally at the transmitter and the receiver, feedback also increases  the capacity of this channel.

The fact that feedback increases the capacity of the energy harvesting channel is indeed surprising. In a classical wireless channel, it is clear that feedback can increase the capacity by allowing the transmitter to learn the state of the channel which is typically available at the receiver. However, in an energy harvesting channel the state of the system (captured by the available energy in the battery) is readily known at the transmitter (but not at the receiver) and communication occurs over a memoryless channel. It is tempting to believe that, since all information regarding the state of the channel is already available at the transmitter, feedback from the receiver will not provide the transmitter with any additional information and therefore will not increase the capacity of this channel. Indeed, it is interesting to note that in his 1956 paper on zero error capacity~\cite{shannon1956zero}, Shannon claims that feedback would not increase the capacity of such channels. Theorem 6 of his paper proves that feedback does not increase the capacity of a discrete memoryless point-to-point channel. His proof is followed by the following interesting comment: 

 ``\textit{It is interesting that the first sentence of Theorem~6 can be generalized readily to channels with memory provided they are of such a nature that the internal state of the channel can be calculated at the transmitting point from the initial state and the sequence of letters
that have been transmitted.}''\footnote{The first sentence of Theorem 6 reads ``\textit{In a memoryless~discrete channel with feedback, the forward capacity is equal to the ordinary capacity C (without feed-back).}''}
\smallbreak
\noindent  Our channel model described in Section~\ref{sec:channel_model} corresponds to a time-invariant finite state channel where the state is computable at the transmitter from the initial state and the transmitted symbol sequence. As such, it provides a counter-example to Shannon's claim.

\section{System Model}
\label{sec:channel_model}

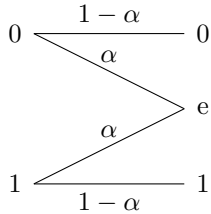
\begin{figure}[!t]
\centering
\begin{tikzpicture}
	\def \h {2};
	
	\draw (0,0) -- node[above] {$\alpha$} (2,\h/2);
	\draw (0,\h) -- node[above] {$\alpha$} (2,\h/2);
	\draw (0,0) -- node[below] {$1-\alpha$} (2,0);
	\draw (0,\h) -- node[above] {$1-\alpha$} (2,\h);
	
	\def \labelpos {0.25};	
	
	\node at (-\labelpos,0) {$1$};
	\node at (2+\labelpos,0) {$1$};
	\node at (-\labelpos,\h) {$0$};
	\node at (2+\labelpos,\h) {$0$};
	\node at (2+\labelpos,\h/2) {$\mathrm{e}$};
\end{tikzpicture}
\caption{Binary erasure channel.}
\label{fig:BEC}
\end{figure}

The energy harvesting binary erasure channel (EH-BEC) depicted in Fig.~\ref{fig:sysmodel} has input alphabet $\mathcal{X}=\{0,1\}$, output alphabet $\mathcal{Y}=\{0,1,\mathrm{e}\}$, and channel transition probabilities given in Fig.~\ref{fig:BEC}.
The transmitter has a battery with finite capacity $B_{\mathrm{max}}=1$, and the input symbol energy at each time slot is constrained by the available energy in the battery.
Let $B_t$ represent the available energy in the battery at time $t$.
The system energy constraints can be described as
\begin{align}
	X_t&\leq B_t,\label{eq:input_constraint}\\
	B_t&=\min\{B_{t-1}-X_{t-1}+E_t,1\},
\end{align}
where $E_t$ is an exogenous process of energy arrivals.
Tutuncuoglu et al.~\cite{tutuncuoglu2013binary,tutuncuoglu2014binary} considered similar binary channels with a unit sized battery, with a noiseless channel and a binary symmetric channel (BSC) instead of the BEC, and i.i.d. Bernoulli energy arrivals.

Here we consider a special case where the energy arrivals $E_t$ are binary and deterministic. In particular, suppose
\[
	E_t=\begin{cases}
		1&,t\text{ odd}\\
		0&,t\text{ even}
	\end{cases}
\]
Therefore, $B_t$ can be written as
\begin{equation}
	B_t=\begin{cases}
		1&,t\text{ odd}\\
		1-X_{t-1}&,t\text{ even}
	\end{cases}
	\label{eq:EH_battery}
\end{equation}

We consider this channel with and without feedback.
An $(M,n)$ code for the EH-BEC without feedback is an encoding function $f$ and a decoding function $g$:
\begin{align}
	f&:\mathcal{M}\to\mathcal{X}^n,\label{eq:enc_func}\\*
	g&:\mathcal{Y}^n\to\mathcal{M}.\label{eq:dec_func}
\end{align}
where $\mathcal{M}=\{1,\ldots,M\}$.
To transmit message $w\in\mathcal{M}$ the transmitter sets $x^n=f(w)$.
The function $f$ must satisfy the energy constraint~\eqref{eq:input_constraint}:
$f_t(w)\leq b_t(f_{t-1}(w))$.
The receiver sets $\hat{W}=g(Y^n)$.

When there is feedback from the receiver to the transmitter, the encoding function~\eqref{eq:enc_func} is changed to
\begin{equation}
	f_t:\mathcal{M}\times\mathcal{Y}^{t-1}\to\mathcal{X},
	\label{eq:fb_enc_func}
\end{equation}
such that $x_t=f_t(w,Y^{t-1})$.
In both cases, the capacity is defined in the usual way as the supremum of all achievable rates.\looseness=-1



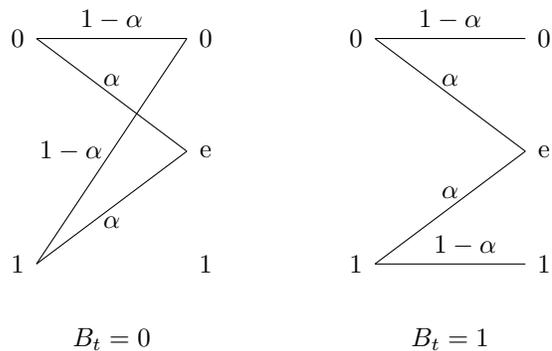
\begin{figure}[!t]
\centering
\begin{tikzpicture}
	\def \labelpos {0.25};	

	\node at (-\labelpos,0) {$1$};
	\node at (2+\labelpos,1.5) {$\mathrm{e}$};
	\node at (2+\labelpos,3) {$0$};
	\node at (-\labelpos,3) {$0$};
	\node at (2+\labelpos,0) {$1$};
	\node at (1,-1) {$B_t=0$};

	\draw (0,0) -- node[above,left] {$1-\alpha$} (2,3);
	\draw (0,0) -- node[below] {$\alpha$} (2,1.5);
	\draw (0,3) -- node[above] {$1-\alpha$} (2,3);
	\draw (0,3) -- node[above] {$\alpha$} (2,1.5);

	\def \r {4.5}

	\draw (\r,0) -- node[above] {$\alpha$} (\r+2,1.5);
	\draw (\r,3) -- node[above] {$\alpha$} (\r+2,1.5);
	\draw (\r,0) -- node[above,pos=0.6] {$1-\alpha$} (\r+2,0);
	\draw (\r,3) -- node[above] {$1-\alpha$} (\r+2,3);
	
	\node at (\r-\labelpos,0) {$1$};
	\node at (\r+2+\labelpos,0) {$1$};
	\node at (\r-\labelpos,3) {$0$};
	\node at (\r+2+\labelpos,3) {$0$};
	\node at (\r+2+\labelpos,1.5) {$\mathrm{e}$};
	\node at (\r+1,-1) {$B_t=1$};
			
\end{tikzpicture}
\caption{Finite state energy harvesting binary erasure channel.}
\label{fig:BEC_states}
\end{figure}

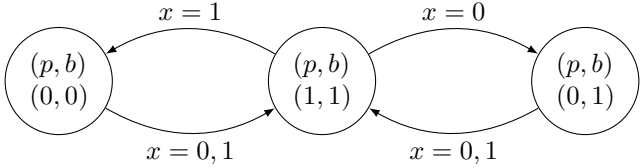
\begin{figure}[!t]
\centering
\begin{tikzpicture}
	\def \tw {2.2em}
	\node[draw,circle,text width=\tw] at (0,0) 
		(11) {$(p,b)$ ${(1,1)}$};
	\node[draw,circle,text width=\tw] at (-3.5,0)
		(00) {$(p,b)$ $(0,0)$};
	\node[draw,circle,text width=\tw] at (3.5,0)
		(01) {$(p,b)$ $(0,1)$};
		
	\def \a {30}
	
	\draw[->,>=latex] (11) to[out=180-\a,in=\a] 
		node[above] {$x=1$} (00);
	\draw[->,>=latex] (00) to[out=-\a,in=180+\a] 
		node[below] {$x=0,1$} (11);
	\draw[->,>=latex] (11) to[out=\a,in=180-\a] 
		node[above] {$x=0$} (01);
	\draw[->,>=latex] (01) to[out=180+\a, in=-\a] 
		node[below] {$x=0,1$} (11);
\end{tikzpicture}
\caption{State diagram of the energy harvesting binary erasure channel.}
\label{fig:state_diagram}
\end{figure}

It is interesting to note that this channel, with or without feedback, is equivalent to a  finite state time-invariant channel where the transmitter can compute the state from the initial state of the channel and the transmitted symbol sequence, satisfying the conditions of Shannon's claim discussed in the earlier section. Let $S_t=(P_t,B_t)$ where $B_t, P_t\in\{0,1\}$ and 
\begin{align*}
	B_{t+1}&=\begin{cases}
		1&,P_t=0\\
		1-X_t&,P_t=1
	\end{cases}\\*
	P_{t+1}&=1-P_t.
\end{align*}
Consider a binary channel with no input constraints, but instead assume that 
when $B_t=1$, the channel behaves as a standard BEC and when $B_t=0$, the channel behaves as a BEC with $X=0$ at its input, regardless of the actual input $X_t$. 
This channel is illustrated in Fig.~\ref{fig:BEC_states}.
Assume the initial state is $s_1=(p_1,b_1)=(1,1)$, and it is known beforehand both at the transmitter and the receiver. Here the state variable $B_t$ corresponds to the battery level in the energy harvesting channel and $P_t\in\{0,1\}$ is a state variable indicating whether the time $t$ is odd or even ($P$ stands for ``parity''). The state diagram is shown in Fig.~\ref{fig:state_diagram}. Note that at odd times, the state always reverts to $s=(1,1)$. At even times, the state is a deterministic function of the past input, therefore it is computable at the transmitter, but unknown at the receiver.
It is easy to see that this time-invariant finite state channel with no input constraints is equivalent to our original EH-BEC, as codes designed for one channel can be easily translated to the other with the same probability of error.

%

\section{Capacity Without Feedback}
\label{sec:capacity_no_fb}

The impact of the energy constraint in \eqref{eq:EH_battery} is to prohibit the transmission of the input $(x_1,x_2)=(1,1)$ over a block of two channel uses starting with an odd channel use. With this additional input constraint, the channel is memoryless over blocks of two channel uses from $X^2$ to $Y^2$. The capacity is then
\begin{align}
	C(\alpha)&=\frac{1}{2}\max_{X^2\neq(1,1)}I(X^2;Y^2)\nonumber\\
	&=\frac{1}{2}\max_{X^2\neq(1,1)}[H(Y^2)-H(Y^2|X^2)]\nonumber\\
	&=\frac{1}{2}\max_{X^2\neq(1,1)}H(Y^2)-h_2(\alpha),
	\label{eq:capacity_no_feedback}
\end{align}
where $h_2(\cdot)$ is the binary entropy function, i.e. $h_2(\alpha)=-\alpha\log_2\alpha-(1-\alpha)\log_2(1-\alpha)$.

To find the optimal input distribution, we first observe that since the channel is memoryless, then by the symmetry and the concavity of the mutual information, the inputs $(0,1)$ and $(1,0)$ must have the same probability, denoted $\pi<0.5$. Then $p(x^2=(0,0))=1-2\pi$.
The entropy of the output can be readily computed, yielding
\begin{equation}
I(X^2;Y^2)=(1-\alpha)^2[h_2(2\pi)+2\pi]
	+2\alpha(1-\alpha)h_2(\pi).
\end{equation}
This is a concave function of $\pi$.
To find the maximum, we take derivative w.r.t. $\pi$ and equate to 0:
\[
	(1-\alpha)^2\left[2\log\frac{1-2\pi}{2\pi}+2\right]
	+2\alpha(1-\alpha)\log\frac{1-\pi}{\pi}
	=0
\]
\begin{align*}
\left(\frac{2\pi}{1-2\pi}\right)\cdot
\left(\frac{\pi}{1-\pi}\right)^{\frac{\alpha}{1-\alpha}}
&=2.
\end{align*}
Denoting $x=\frac{\pi}{1-\pi}$, we get
\[
	x^{1/(1-\alpha)}+x-1=0.
\]

This can be solved numerically for any value of $0<\alpha<1$.
Specifically, for $\alpha=0.5$ we can solve analytically to obtain $\pi=(3-\sqrt{5})/2\approx 0.382$.
Substituting in the expression for capacity, we have
\begin{align*}
	C(0.5)
	&=\frac{1}{8}[h_2(3-\sqrt{5})
	+2h_2((3-\sqrt{5})/2)+3-\sqrt{5}]\\
	&=0.4339.
\end{align*}

\section{Capacity With Feedback}
\label{sec:capacity_fb}

Consider now the channel defined in Section~\ref{sec:channel_model} with feedback.
Looking at blocks of size 2, the channel is memoryless over different blocks, but with \emph{in-block memory}.
The capacity of this channel is given by
\begin{equation}
	C_\text{fb}(\alpha)=
	\frac{1}{2}\max_{p(x^2\|y_1)}
	I(X^2\to Y^2),
\label{eq:Cfb_directed_information}
\end{equation}
where $I(X^2\to Y^2)$ is directed information and $p(x^2\|y_1)=p(x_1)\cdot p(x_2|x_1,y_1)$ is the causal conditioning input distribution with the additional constraint
\[ p({x_2=1}|{x_1=1},y_1)=0 
\quad\forall y_1.\]
imposed by energy harvesting model. This result has been established for a far more general case in~\cite{kramer2014information} (specifically, we apply here Theorem~2 and eq.~(48) therein).


%

Let 
\[ p(x_1=1)=p_1,\]
\begin{align*}
p(x_2=1|x_1=0,y_1=0)&=p_{20},\\*
p(x_2=1|x_1=0,y_1=\text{e})&=p_{2e},
\end{align*}
where $0\leq p_1,p_{20},p_{2e}\leq 1$.
Then the directed information in~\eqref{eq:Cfb_directed_information} can be written as
\begin{align*}
I(X^2\to Y^2)
&=I(X_1;Y_1)+I(X^2;Y_2|Y_1)\\
&=H(Y_1)+H(Y_2|Y_1)-2h_2(\alpha),
\end{align*}
where 
\begin{align*}
	H(Y_1)&=h_2(\alpha)+(1-\alpha)h_2(p_1),\\
	H(Y_2|Y_1)&=(1-p_1)(1-\alpha)H(Y_2|Y_1=0)\\*
		&\quad+p_1(1-\alpha)H(Y_2|Y_1=1)
		+\alpha H(Y_2|Y_1=\text{e}).
\end{align*}
Clearly, $Y_1=1$ implies $X_1=1$, which in turn implies ${X_2=0}$. Therefore $H(Y_2|Y_1=1)=h_2(\alpha)$.
When $Y_1=0$, the input is necessarily $X_1=0$, then the input $X_2$ is $\text{Bernoulli}(p_{20})$, which yields 
\[
H(Y_2|Y_1=0)=h_2(\alpha)+(1-\alpha)h_2(p_{20}).
\]
Finally, when $Y_1=\text{e}$, we have $X_2=1$ w.p. $p_{2e}$ only if $X_1=0$, and $X_2=0$ otherwise. Therefore $X_2\sim\text{Bernoulli}{\big(p_{2e}(1-p_1)\big)}$, giving
\[
H(Y_2|Y_1=\text{e})=h_2(\alpha)+(1-\alpha)h_2\big(p_{2e}(1-p_1)\big).
\]
Summing up all terms, we get
\begin{align*}
	I(X^2\to Y^2)&=(1-\alpha)\big[h_2(p_1)
		+(1-\alpha)(1-p_1)h_2(p_{20})\\*
		&\hspace{5em}
		+\alpha h_2\big(p_{2e}(1-p_1)\big)\big].
\end{align*}

This can be maximized by choosing $p_{20}=0.5$ and 
$p_{2e}=\min\{\frac{1}{2(1-p_1)},1\}$.
Further optimization over $p_1\in[0,1]$ yields
\[
	p_1=\frac{1}{1+2^{1-\alpha}}.
\]
We finally get
\[
	C_\text{fb}(\alpha)=
	\frac{1-\alpha}{2}\big[\log(1+2^{1-\alpha})+\alpha\big].
\]

\begin{figure}[!t]
\begin{center}
\includegraphics[width=2.5in]{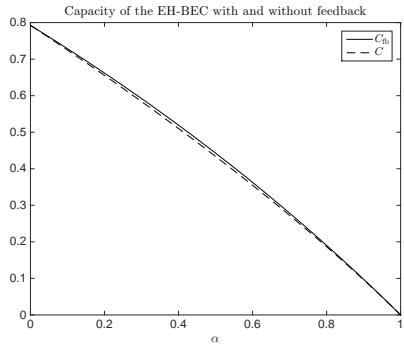}
\end{center}
\caption{Capacity of the EH-BEC with periodical recharges, with and without feedback.}
\label{fig:capacities}
\end{figure}

For $\alpha=0.5$, we get
\[
	C_\text{fb}(0.5)=
	 0.4429 > 0.4339 = C(0.5).
\]
For all other values of $0\leq\alpha\leq 1$, the capacities with and without feedback are plotted in Fig.~\ref{fig:capacities}.

\section{Why does Feedback Help?}
\label{sec:equivalent_state_dependent_model}

In this section, we will try to illustrate the intuition behind the usefulness of feedback in this scenario. Recall that the state of the battery is
\[
B_t=\begin{cases}
1&,t\ \mathrm{odd}\\
1-X_{t-1}&,t\ \mathrm{even}
\end{cases}
\]
We focus on even times: the transmitter knows the current state of the channel $B_t$, and the receiver has a noisy estimate of it $\tilde{B}_t=Y_{t-1}$. (Note that $Y_{t-1}$ is the output of a BEC with input $X_{t-1}=1-B_t$.) With feedback, the transmitter not only knows the true state of the channel $B_t$ but also its noisy estimate at the receiver  $\tilde{B}_t$. 

\begin{figure}[!th]
\centering
\begin{tikzpicture}
	\node[draw,rectangle] (Enc) at (0,0) {Enc.};
	\node[draw,rectangle] (Channel) at (3,0) {Channel};
	\node at (3,0.5) {BEC};
	\node[draw,rectangle] (Dec) at (6,0) {Dec.};
	\node[draw,rectangle] (StateGen) at (3,-3) 
		{State Gen.};

	\def\r{0.4};	
	
	\node[draw=black,circle,inner sep=0pt,minimum size=1mm,fill=black] (Sbranch) at (3-\r,-2) {};
	\node[draw=black,circle,inner sep=0pt,minimum size=1mm,fill=black] (Stildebranch) at (3+\r,-1.25) {};
	\node (Semicircle) at (3-\r,-1.25) {};
	
	\tikzstyle{every path}=[draw,->]
	\path (Enc) -- node[above] {$X_t$} (Channel);
	\path (Channel) -- node[above] {$Y_t$} (Dec);
	\path ($(StateGen.north) - (\r,0)$) -- ($(Channel.south) - (\r,0)$);
	\path (Sbranch) -| node[below,pos=0.05] {$S_t$} ($(Enc.south) - (0.2,0)$);
	\path ($(StateGen.north) + (\r,0)$) -- node[right,pos=0.2] {$\tilde{S}_t$} (Stildebranch) -| (Dec);
	\path[dashed] (Stildebranch) -| ($(Enc.south) + (0.2,0)$);
\end{tikzpicture}
\caption{Equivalent channel models. The dashed line corresponds to the channel equivalent to the case with feedback.}
\label{fig:equivalent_channels}
\end{figure}
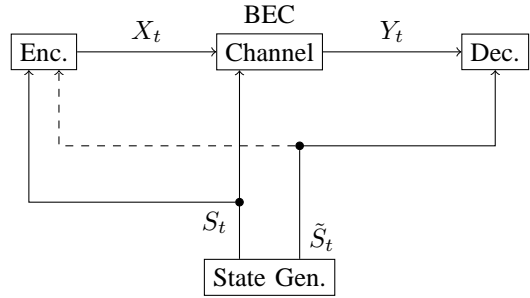

The question of whether feedback can help to increase the capacity of this channel is then related to the following question: Consider a channel with i.i.d. states $S_t$ known causally at the transmitter. Assume the receiver observes a noisy version of the state $\tilde{S}_t$. Can the capacity be increased if the transmitter knew the receiver's noisy estimate of the state $\tilde{S}_t$ in addition to knowing the actual state of the channel? See Fig~\ref{fig:equivalent_channels}. 
The capacity when the transmitter observes only $S_t$ is given by
\begin{align*}
C&=\max_{p(u)}I(U;Y,\tilde{S})
=\max_{p(u)}I(U;Y|\tilde{S})
&,U:\mathcal{S}\to \mathcal{X}.
\end{align*}
When transmitter also observes $\tilde{S}_t$, the capacity is 
\begin{align*}
C_{\mathrm{fb}}&=\max_{p(u|\tilde{s})}I(U;Y|\tilde{S})
&,U:\mathcal{S}\to \mathcal{X}.
\end{align*}
The increase in capacity follows from allowing $U$ to depend on $\tilde{S}$.


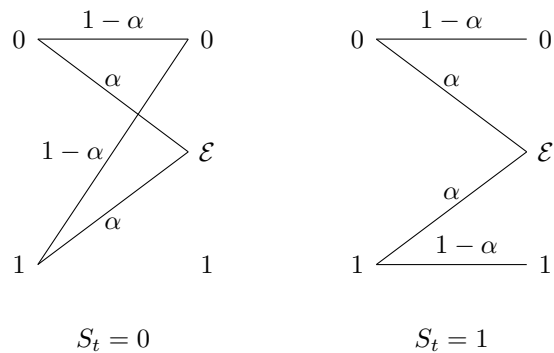
\begin{figure}[!th]
\centering
\begin{tikzpicture}
	\def \labelpos {0.25};	

	\node at (-\labelpos,0) {$1$};
	\node at (2+\labelpos,1.5) {$\mathcal{E}$};
	\node at (2+\labelpos,3) {$0$};
	\node at (-\labelpos,3) {$0$};
	\node at (2+\labelpos,0) {$1$};
	\node at (1,-1) {$S_t=0$};

	\draw (0,0) -- node[above,left] {$1-\alpha$} (2,3);
	\draw (0,0) -- node[below] {$\alpha$} (2,1.5);
	\draw (0,3) -- node[above] {$1-\alpha$} (2,3);
	\draw (0,3) -- node[above] {$\alpha$} (2,1.5);

	\def \r {4.5}

	\draw (\r,0) -- node[above] {$\alpha$} (\r+2,1.5);
	\draw (\r,3) -- node[above] {$\alpha$} (\r+2,1.5);
	\draw (\r,0) -- node[above,pos=0.6] {$1-\alpha$} (\r+2,0);
	\draw (\r,3) -- node[above] {$1-\alpha$} (\r+2,3);
	
	\node at (\r-\labelpos,0) {$1$};
	\node at (\r+2+\labelpos,0) {$1$};
	\node at (\r-\labelpos,3) {$0$};
	\node at (\r+2+\labelpos,3) {$0$};
	\node at (\r+2+\labelpos,1.5) {$\mathcal{E}$};
	\node at (\r+1,-1) {$S_t=1$};
			
\end{tikzpicture}
\caption{Equivalent channel with i.i.d. states.}
\label{fig:iid_states}
\end{figure}

To illustrate that the second capacity can be strictly larger than the first assume $S_t=1$ w.p. $p$ and $0$ w.p $1-p$. $\tilde{S}_t$ is given as the output of a $\mathrm{BEC}(\alpha)$ with $S_t$ as its input. The channel transition probabilities depend on $S_t$ as in 
Fig.~\ref{fig:iid_states} (note that this is exactly the same as Fig.~\ref{fig:BEC_states}, with $B_t$ replaced by $S_t$).\footnote{
This channel does not exactly correspond to our original EH-BEC and has  different capacity. We use it to illustrate how feedback can be useful to increase the capacity of the EH-BEC, rather than providing a direct equivalence.} The capacities in the two cases above can be explicitly computed:
\begin{align*}
C&=(1-\alpha)\max_{0\leq r\leq 1}
\big[p(1-\alpha)h_2(r)+\alpha(h_2(pr)-rh_2(p))\big]\\
C_{\mathrm{fb}}&=(1-\alpha)\max_{0\leq r\leq 1}
\big[p(1-\alpha)+\alpha(h_2(pr)-rh_2(p))\big].
\end{align*}
We can see that $C_{\mathrm{fb}}\geq C$ with equality iff $r=1/2$, which is true only when $p=0$ or $p=1$, or when $\alpha=0$ or $\alpha=1$.

\section{Random Energy Arrivals}
\label{sec:random_energy_arrivals}

We showed that feedback can increase capacity when the energy arrivals are deterministic with period 2.
However, the model usually studied in the literature involves i.i.d. energy arrivals (see e.g.~\cite{tutuncuoglu2013binary,tutuncuoglu2014binary}).
We will show that feedback can help in this case as well, at least when  noncausal observations of the energy arrivals are available at the transmitter and the receiver.

Consider the model presented in Section~\ref{sec:channel_model}, with the following modifications:
the energy arrivals $E_t$ are now i.i.d. $\text{Bernoulli}(p)$ RVs known noncausally to both the transmitter and the receiver.
The encoder and decoder functions~\eqref{eq:enc_func} and~\eqref{eq:dec_func} are now modified to
\begin{align}
	f&:\mathcal{M}\times\mathcal{E}^n
		\to\mathcal{X}^n,\\*
	g&:\mathcal{Y}^n\times\mathcal{E}^n
		\to\mathcal{M},
\end{align}
where $\mathcal{E}=\{0,1\}$ is the alphabet of $E_t$.
Similarly to~\eqref{eq:fb_enc_func}, when there is feedback the encoding function becomes
\begin{equation}
	f_t:\mathcal{M}\times\mathcal{E}^n\times
		\mathcal{Y}^{t-1}\to\mathcal{X}.
\end{equation}

We prove the following theorem in Appendix~\ref{sec:noncausal_capacity_proof}.
\begin{theorem}
\label{thm:noncausal_capacity}
	The capacity of the EH-BEC with i.i.d. $\text{Bernoulli}(p)$ energy arrivals and noncausal energy arrival information, with and without feedback, is given by
	\begin{align}
	C&=\sum_{k=1}^{\infty}p^2(1-p)^{k-1}
		\max_{\substack{p(x^k):\\ \sum_{i=1}^{k}X_i\leq1}}
		H(Y^k)-h_2(\alpha),
		\label{eq:C_noncausal}\\
	C_{\mathrm{fb}}&=\sum_{k=1}^{\infty}p^2(1-p)^{k-1}
		\max_{\substack{p(x^k\|y^{k-1}):\\ 
			\sum_{i=1}^{k}X_i\leq1}}
		H(Y^k)-h_2(\alpha),
		\label{eq:Cfb_noncausal}
	\end{align}
where $p(x^k\|y^{k-1})=\prod_{i=1}^{n}p(x_i|x^{i-1},y^{i-1})$ is a causally conditioned input distribution, and in both cases the maximization is over all input distributions with support $\sum_{i=1}^{k}X_i\leq 1$ a.s., i.e. only input sequences with at most one 1.
\end{theorem}

Observe that for every $k$ we have
\[
	\max_{\substack{p(x^k\|y^{k-1}):\\ 
			\sum_{i=1}^{k}X_i\leq1}}
		H(Y^k)
	\geq \max_{\substack{p(x^k):\\ \sum_{i=1}^{k}X_i\leq1}}
		H(Y^k),
\]
and the results of Sections~\ref{sec:capacity_no_fb} and~\ref{sec:capacity_fb} imply that the inequality is strict for $k=2$ and $\alpha=0.5$, that is
\[
	\max_{\substack{p(x^2\|y_1):\\ X_1+X_2\leq1}}
		H(Y^2)
	> \max_{\substack{p(x^2):\\ X_1+X_2\leq1}}
		H(Y^2).
\]
Therefore, we conclude that feedback can strictly increase capacity for i.i.d. energy arrivals.

\appendices

\section{Proof of Theorem~\ref{thm:noncausal_capacity}}
\label{sec:noncausal_capacity_proof}

First we note that $C\leq C_{\mathrm{fb}}\leq 1-\alpha$, and both $C$ and $C_{\mathrm{fb}}$ are limits of increasing sequences, so convergence is guaranteed.
We prove Theorem~\ref{thm:noncausal_capacity} for the case with feedback; the proof without feedback follows exactly the same lines.

\subsection{Achievability}
Without loss of generality, we assume the initial battery state is $B_0=1$, or, equivalently, that $E_1=1$ w.p. 1. 
Fix $N$ and maximizing distributions $\{p(x^{k}\|y^{k-1})\}_{k=1}^{N}$ in~\eqref{eq:Cfb_noncausal}.
Divide the message into $N$ messages, such that $R=\sum_{k=1}^{N}R_{k}$.
Upon observing $e^n$, the transmitter and receiver divide the transmission into \emph{epochs}, where an epoch refers to the time between two consecutive energy arrivals. More precisely, $e^n$ can be mapped to a sequence of integers $(\ell_1,\ell_2,\ldots,\ell_m)$, where $m(e^n)=\sum_{t=1}^{n}e_t$ is the number of energy arrivals, $\ell_i(e^n)$ is the time between the $i$-th and the $(i+1)$-th energy arrivals, and we let $\ell_m=n-\sum_{i=1}^{m-1}\ell_i$. Each epoch can be considered as a super-symbol and the epoch length can be thought of as the random state of the channel determining the size of the inputted super-symbol. For this super-channel with states we use the multiplexing technique in~\cite[Section~7.4.1]{elgamal2011network} to communicate a codeword of rate $R_k$ over each state $k$. (Since each $\ell_i$ can take any value between $1$ and $n$, we treat all values greater than or equal to $N$ as the state $k=N$.) For each state $k$, we generate a codeword where each super-symbol $x^k$ is generated according to the pmf $p(x^k\|y^{k-1})$.
Note that this guarantees the energy constraint~\eqref{eq:input_constraint} is satisfied.
For decoding the codeword corresponding to state $k$, we use the technique in~\cite{kim2008coding},\cite[Section~17.6.3]{elgamal2011network}.
Roughly, the subcodeword formed by the $j$-th symbol inside the super-symbol $x^k$ is decoded separately for $1\leq j\leq k$ by treating the earlier decoded subcodewords and the corresponding channel outputs as side information. Thus, for sub-block $j$ we can achieve rate $I(X_j;Y_j^k|X^{j-1},Y^{j-1})$.
The achievable rate for state $k$ is then given by
\begin{align*}
R_k&=\sum_{j=1}^{k}I(X_j;Y_j^k|X^{j-1},Y^{j-1})\\
&=I(X^k\to Y^k)\\
&\stackrel{\text{(a)}}{=}H(Y^k)-kh_2(\alpha),
\end{align*}
where (a) is due to the memorylessness of the channel and because $H(Y_i|X_i)=h_2(\alpha)$.

For a sequence $e^n$ and $1\leq k\leq N$, the empirical distribution of the states, or epoch lengths, is 
\[
\pi(k|e^n)=\frac{1}{n}\sum_{i=1}^{m(e^n)}
\mathbf{1}\{\min(\ell_i(e^n),N)=k\},
\]
where $\mathbf{1}\{\cdot\}$ is the indicator function, and $m(e^n)$ and $\ell_i(e^n)$, $i=1,\ldots,m(e^n)$, have been defined above.
Note that this is not a legitimate probability distribution, as it does not sum to 1.
Nevertheless, by the strong law of large numbers for regenerative processes:
\begin{align*}
\pi(k|E^n)
&=\frac{m(E^n)}{n}\cdot\frac{1}{m(E^n)}\sum_{i=1}^{m(E^n)}
\mathbf{1}\{\min(\ell_i(E^n),N)=k\}\\
&\to p\cdot q(k)\text{ a.s. as }n\to\infty,
\end{align*}
where $q(k)$ is a probability distribution, defined as
\[
q(k)=\begin{cases}
	p(1-p)^{k-1}&,1\leq k\leq N-1\\
	(1-p)^{N-1}&,k=N
\end{cases}
\]
We define the following $\epsilon$-typical set for $e^n$:
\begin{align*}
\mathcal{T}_{\epsilon}^{(n)}=
\big\{e^n:\ &\big|\pi(k|e^n)-pq(k)\big|
	\leq\epsilon pq(k),\ 
	\forall 1\leq k\leq N
\big\}.
\end{align*}
Hence $\Pr\{E^n\in\mathcal{T}_{\epsilon}^{(n)}\}\to1$.

Now, assuming $e^n\in\mathcal{T}_{\epsilon}^{(n)}$, there are $n_k\geq n(1-\epsilon)pq(k)$ symbols transmitted in state $k$.
The achievable rate is then
\begin{align*}
R&=\sum_{k=1}^{N}(1-\epsilon)pq(k)R_k\\
&\geq(1-\epsilon)\sum_{k=1}^{N}p^2(1-p)^{k-1}
	[H(Y^k)-kh_2(\alpha)].
\end{align*}
This is a lower bound to capacity for every $\epsilon>0$ and $N\geq1$, therefore we can take $\epsilon\to0$ and $N\to\infty$ to obtain
\[
C_{\mathrm{fb}}\geq\sum_{k=1}^{\infty}p^2(1-p)^{k-1}H(Y^k)-h_2(\alpha).
\]

\subsection{Converse}
By Fano's inequality:
\begin{align*}
nR-n\epsilon_n
&\leq I(W;Y^n|E^n)\\
&= \sum_{t=1}^{n}I(W;Y_t|Y^{t-1},E^n)\\
&= \sum_{t=1}^{n}I(X_t;Y_t|Y^{t-1},E^n)\\
&= \sum_{t=1}^{n}[H(Y_t|Y^{t-1},E^n)-h_2(\alpha)]\\
&= H(Y^n|E^n)-nh_2(\alpha)\\
&=\sum_{e^n}p(e^n)H(Y^n|E^n=e^n)-nh_2(\alpha).
\end{align*}
Recall the definition of $m(e^n)$ and $\ell_i(e^n)$, $i=1,\ldots,m(e^n)$, as before.
We further define $t_i(e^n)$, $i=1,\ldots,m(e^n)$, as the energy arrival times, i.e. the times for which $e_t=1$, or $t_i=1+\sum_{j=1}^{i-1}\ell_j(e^n)$ (where again we assume $E_t=1$ w.p. 1).
Then we can further upper-bound the rate as
\begin{align*}
nR-n\epsilon_n
&\leq \sum_{e^n}p(e^n)\sum_{i=1}^{m(e^n)}
	H(Y_{t_i}^{t_{i+1}-1})-nh_2(\alpha)\\
&\leq\sum_{e^n}p(e^n)\sum_{i=1}^{m(e^n)}
	C(\ell_i),
\end{align*}
where
\[
C(k)\triangleq \max_{\substack{p(x^k\|y^{k-1}):\\ 
\sum_{i=1}^{k}X_i\leq 1}}H(Y^k)-kh_2(\alpha).
\]
Taking $n\to\infty$, we get
\[
C_{\mathrm{fb}}\leq\liminf_{n\to\infty}\frac{1}{n}
\mathbb{E}\left[\sum_{i=1}^{m(E^n)}C\big(\ell_i(E^n)\big)\right],
\]
where the expectation is over the RV $E^n$.
By the strong law of large numbers for regenerative processes:
\[
\frac{m(E^n)}{n}\cdot\frac{1}{m(E^n)}
\sum_{i=1}^{m(E^n)}C\big(\ell_i(E^n)\big)
\to p\cdot\mathbb{E}[C(L)]
\quad\text{a.s.},
\]
where $L$ is a geometric RV with parameter $p$.
Moreover, since $C(k)\leq k(1-\alpha)$ and
$n=\sum_{i=1}^{m(e^n)}\ell_i(e^n)$ for any $e^n$:
\[
\frac{1}{n}\sum_{i=1}^{m(E^n)}C\big(\ell_i(E^n)\big)
\leq(1-\alpha)
\quad\text{w.p. 1},
\]
Therefore, by bounded convergence, the above limit converges and it is given by
\begin{align*}
C_{\mathrm{fb}}&\leq p\cdot\mathbb{E}[C(L)]\\
&=p\sum_{k=1}^{\infty}p(1-p)^{k-1}C(k)\\
&=\sum_{k=1}^{\infty}p^2(1-p)^{k-1}
	\max_{\substack{p(x^k\|y^{k-1}):\\ 
	\sum_{i=1}^{k}X_i\leq 1}}H(Y^k)-h_2(\alpha),
\end{align*}
which completes the proof.

\bibliographystyle{IEEEtran}
\bibliography{IEEEabrv,feedback_BEC}

\end{document}